\documentclass[aps,prx,twocolumn,amsmath,amssymb,nofootinbib,superscriptaddress,floatfix,reprint,longbibliography]{revtex4-2}
\usepackage[dvips]{graphicx}
\usepackage{latexsym}
\usepackage{amsmath}
\usepackage{amsfonts}
\usepackage{amssymb}
\usepackage{bm}
\usepackage{color}
\usepackage{txfonts}
\usepackage{float}
\usepackage{url}
\usepackage[colorlinks=true, urlcolor=blue, linkcolor=blue, citecolor=blue, pdftex]{hyperref}
\usepackage{ulem}
\usepackage{lineno}

\begin{document}

\title {Spin-triplet pair density wave superconductors
}
\author{Yi Zhang}
\email{zhangyi821@shu.edu.cn}
\affiliation{Department of Physics and Institute for Quantum Science and Technology, Shanghai University, Shanghai 200444, China}
\affiliation{Shanghai Key Laboratory of High Temperature Superconductors and International Center of Quantum and Molecular Structures, Shanghai University, Shanghai 200444, China}

\author{Ziqiang Wang}
\email{ziqiang.wang@bc.edu}
\affiliation{Department of Physics, Boston College, Chestnut Hill, MA 02467, USA}

\date{\today}

\begin{abstract}
Recent experiments have shown that the nonzero center of mass momentum pair density wave (PDW) is a widespread phenomenon observed over different superconducting materials. However, concrete theoretical model realizations of the PDW order have remained elusive. Here, we study a
one-dimensional model with nearest-neighbor pairing attraction, i.e. a spinful Kitaev chain, under generic spin-orbit couplings such that the spin-rotation symmetry is fully broken. The most general superconducting order parameter is described by a spatial dependent $\mathbf{d}_i$-vector. 
We show that a spin-triplet pair density wave (t-PDW) emerges in the ground state and occupies a large part of the phase diagram. The $\mathbf{d_i}$-vector of the t-PDW rotates with a pitch $Q_{\rm pdw}$ along the chain and spans an ellipsoid. 
The pure t-PDW is fully gapped and realizes a class-DIII topological superconductor, characterized by two Majorana zero modes localized at each end of the chain and protected by time-reversal symmetry.
Our findings reveal unprecedented insights into the exotic pure PDW superconductor and provide a possible explanation for the one-dimensional PDW detected along domain walls in monolayer iron-based superconductor Fe(Te,Se) and potentially realizable using other quantum structures in unconventional superconductors.
\end{abstract}
\maketitle

\noindent {\large\textbf{Introduction}}\\
In the conventional BCS theory of superconductors, electrons occupying time-reversed quantum states pair together to form Cooper pairs with zero center-of-mass momentum, giving rise to a uniform superconducting order parameter that respects translation symmetry~\cite{bcs}.  
However, in a magnetic field that breaks the time-reversal symmetry, it was proposed that the Cooper pairs can acquire a finite center-of-mass momentum and give rise to the Fulde–Ferrell–Larkin–Ovchinnikov (FFLO) superconductor \cite{FF,LO}
with a spatially modulated superconducting (SC) order parameter.

The FFLO states can be further divided into FF and LL states.
The Cooper pairs in the FF state carry a single nonzero momentum $\mathbf{q}$, such that the SC order parameter breaks time-reversal symmetry and varies in space as $\Delta(\mathbf{r})=\Delta_{\mathbf{q}}e^{i\mathbf{q}\cdot\mathbf{r}}$.
This helical superconductor has been extensively studied in noncentrosymmetric systems with spin-orbit coupling (SOC) and time-reversal symmetry-breaking (TRSB) fields~\cite{Smidman_2017,gorkov_2001,victor_2002, AGTERBERG200313,dim_2003,kaur_2005,kaur_2007,dim_2007,sam_2008,yanase_2008,mich_2012,sek_2013,houzet_2015,yuan_2019}.
It is believed to play an important role in realizing the diode effect in SC systems~\cite{fu_2022,yanse, KFF}.
The LO state, on the other hand, is described by a SC order parameter containing at least two nonzero momenta i.e., $\pm \mathbf{q}$, and varies in space according to $\Delta(\mathbf{r})=\Delta_{\mathbf{q}}\cos{(\mathbf{q}\cdot\mathbf{r})}$.
This inhomogeneous SC state does not require broken time-reversal symmetry. Its generalization in the absence of an external magnetic field is more commonly referred to as the pair density wave (PDW) state~\cite{review_2020}.

The PDW state has been proposed to exist in the high-$T_c$ cuprate superconductors~\cite{cup1,cup2,cup3,cup4,cup5,cup6,cup7,cup8,cup9,cup10,chen_2004,lee_2014} and the  has attracted intense research interest~\cite{erez_2010,jae_2012,jordan_2019,egor_2019,yahui_2022,jiang_2023,feng_2023,xu_2019,cheng_2021,huang_2022,hong_2023,yifang_2023,sen_2022,himeda_2002,marcin_2007,Aperis_2008,Yang_2009,loder_2011,cho_2012,sato_2015,yuxuan_2015,jian_2015,jonatan_2017,jonatan_2018,han_2020,Chakraborty_2021,setty_2023,jin_2022,miao_2016,han_2022,coleman_2022,daniel_2023,shaffer_2023,wu_2023_1,wu_2023_2,tilman_2023,guodong_2023,chandan_2023,guo_2024,huang_2023,huang_2024}.
Moreover, the PDW state is also studied in the spin-triplet pairing phase of He-3 superfluid~\cite{he3_1,he3_2,he3_7,he3_3,he3_5,he3_6}.
More recently, evidence for PDW formation has been reported in iron-based superconductors~\cite{zhao_2023,liu_2023}, heavy-fermion superconductors~\cite{gu_2023,aish_2023}, and kagome superconductors~\cite{hui_2023,deng_2024}. 
Despite these exciting developments, concrete theoretical model realizations of the PDW states beyond phenomenological Landau free-energy descriptions have remained very challenging.

Motivated by the recent experimental progress, especially the detection of the one-dimensional PDW modulations at the domain walls in
the iron-based superconductor Fe(Te,Se) film~\cite{liu_2023}, 
we propose and study a one-dimensional (1D) model with nearest-neighbor pairing attraction and SOCs such that the spin-rotation symmetry is fully broken. 
This simple and concrete model can be thought as a generalization of the spinless Kitaev model to the spinful case in the presence of both Rashba ($\alpha_R$) and Dresselhaus ($\alpha_D$) SOC which couples the two spin species and maintains time-reversal symmetry.  
Since the spin rotation symmetry is fully broken, the pairing interaction can be parametrized independently by attractions in the
equal-spin ($V_1$) and opposite-spin ($V_2$) channels. The nature of the SC states can thus be explored by varying the ratio of $r_V=V_2/V_1$ and $r_{\rm so}=\alpha_D/\alpha_R$.

\begin{figure}
	\begin{center}
        \includegraphics[width=3.4in]{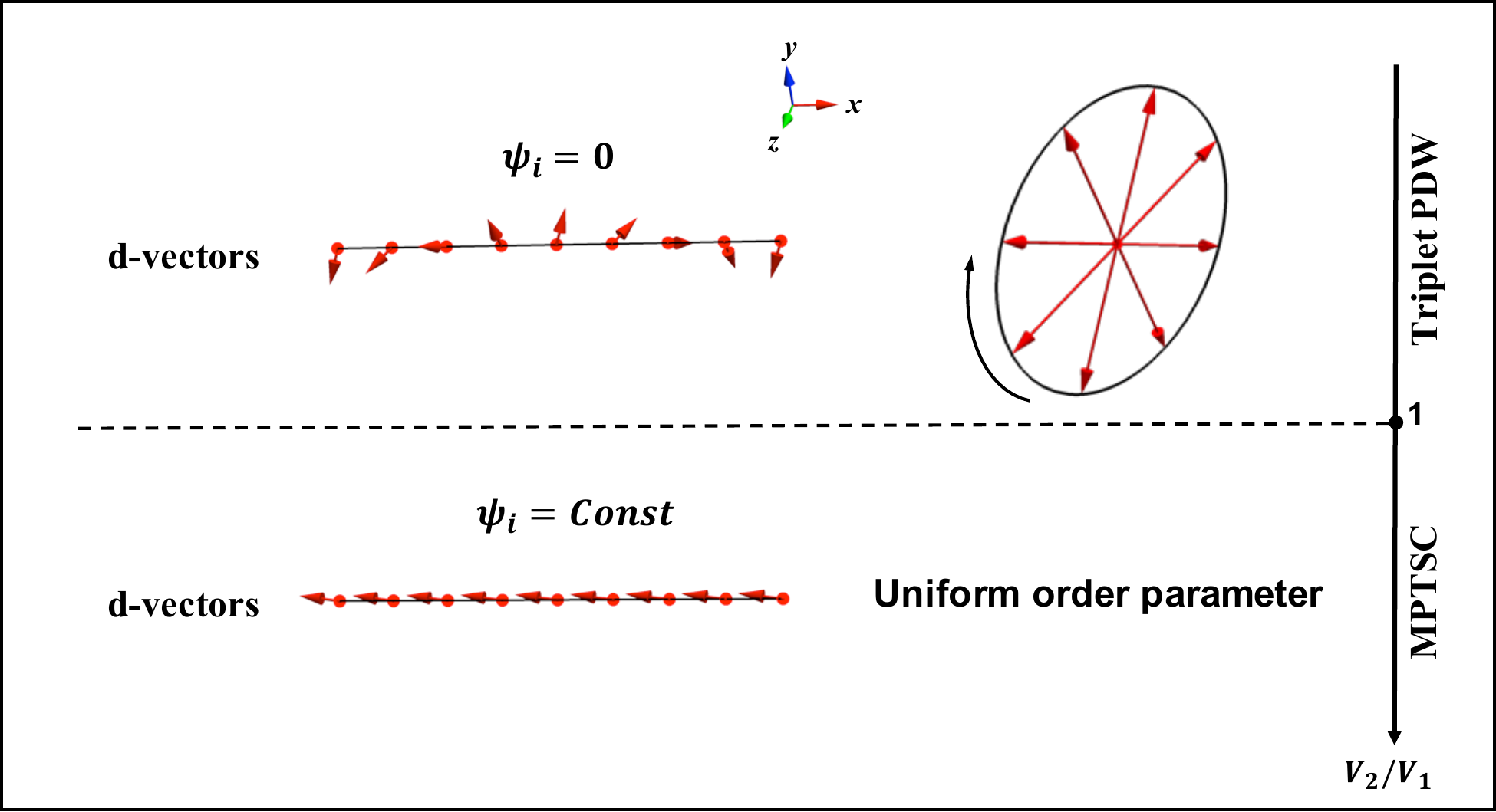}
	\caption{\textbf{Spin-triplet pair density wave (t-PDW) state.}   Schematic phase diagram as the ratio of the attraction $r_V=\frac{V_2}{V_1}$ is tuned for a small value of the ratio of spin-orbit coupling $r_{\rm so}=\frac{\alpha_D}{\alpha_R}$. The phase transition between the t-PDW state and the uniform mixed-parity topological superconductor (MPTSC) state occurs at the transition point with $V_1=V_2$. The t-PDW state is a pure spin-triplet pairing state that can be described by a spatial-dependent imaginary $d$-vector. Its imaginary part is depicted by red arrows showing the spatial modulation. The $d$-vector revolves around the origin in an elliptical orbit in 3D space within a period of the t-PDW state. The uniform order parameters describe the MPTSC, i.e. both $\psi$ and $\textbf{d}$ are uniform in real space.
			\label{fig:pd_gen}}
	\end{center}
	\vskip-0.5cm
\end{figure}

Due to the broken inversion symmetry, the pairing state is in general of a mixed parity type~\cite{Fri_2004}, with the spin-triplet sector specified by the $d$-vector~\cite{review_d}. The most general SC state can thus be described by a spatial-dependent pairing order parameter (Methods)
\begin{equation}
	\Delta_{i,\sigma\sigma^{\prime}}=\left[ (\psi_i \sigma_0 + \mathbf{d}_i\cdot\pmb{\sigma}) i\sigma_y \right]_{\sigma\sigma^{\prime}}
 \label{eq:dv}
\end{equation}
where $\psi_i$ and $\mathbf{d}_i$ correspond to the spin-singlet and the spin-triplet components at site-$i$, respectively.
We perform self-consistent mean-field theory calculations and explore the phase diagram as a function of the SOC ratio $r_{\rm so}$ and the ratio of opposite and equal-spin attractions $r_V$.
We find that a time-reversal symmetric, spin-triplet pair density wave (t-PDW) superconductor emerges in the ground state and occupies a large part of the phase diagram, which also contains a uniform mixed-parity (MP) superconductor. For a fixed SOC ratio $r_{\rm so}$, the phase diagram is shown schematically shown in Fig.~\ref{fig:pd_gen}, displaying a transition from the uniform MP topological superconductor (TSC) for $V_2>V_1$ to a t-PDW TSC for $V_1>V_2$. In the t-PDW phase, the spin-singlet component vanishes, while the
$\mathbf{d_i}$-vector rotates with a pitch $Q_{\rm pdw}$ along the chain and spans an ellipsoid (Fig.~\ref{fig:pd_gen}). We will show that this pure t-PDW is fully-gapped and belongs to the class-DIII TSC with two Majorana zero modes localized at each end of the chain and protected by time-reversal symmetry. Moreover, we study the evolution of the phase diagram with the chemical potential and the properties of the topological phase transitions. Our simple and concrete model allows us to address the critical question of how the pairing interaction and SOC work together to overcome the kinetic energy of the pairs with nonzero center of mass momentum, such that the t-PDW can emerge in the ground state.
The mean-field theory is formulated at zero temperature, where quasi-long-range order can exist with power-law correlations in 1D quantum systems when fluctuations are included beyond the mean-field theory.  \\

\noindent {\large\textbf{Results}}\\
\noindent \textbf{Model}\\
The Hamiltonian of our 1D model of a spin-orbit coupled chain with nearest neighbor (NN) attraction is given by
\begin{equation}
 \begin{split}
	\hat{H}&=\sum_{i\sigma\sigma^{\prime}} c_{i\sigma}^{\dagger} \left(-t\sigma_0+i\alpha_R\sigma_z +i\alpha_D\sigma_x \right)_{\sigma\sigma^{\prime}} c_{i+1\sigma^{\prime}} +h.c.
 \\
 &-V_1\sum_{i\sigma}n_{i\sigma}n_{i+1\sigma}-V_2\sum_{i\sigma}n_{i\sigma}n_{i+1\bar\sigma}
 \end{split}
 \label{eq:ham}
\end{equation}
where $t$ is the NN hopping parameter, $\alpha_R$, and $\alpha_D$ describe the NN Rashba and Dresselhaus SOC, and $V_1$ and $V_2$ are the attractions between equal-spin and opposite-spin charge densities on the NN sites, respectively.

\begin{figure}
	\begin{center}
        \includegraphics[width=3.4in]{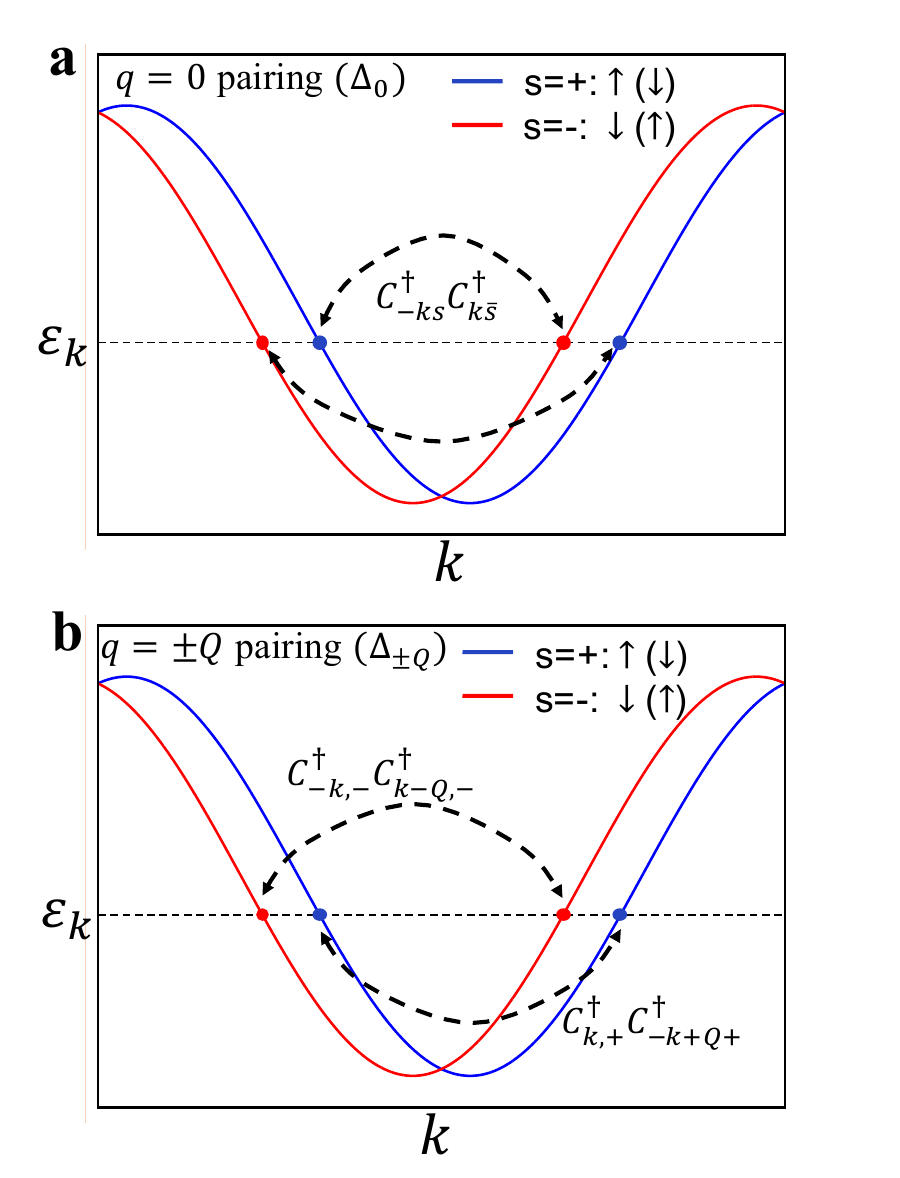}
	\caption{\textbf{Schematic band structure of the spin-orbit coupled chain, showing different pairing channels across the Fermi points.} \textbf{a} The inter-branch pairing leads to the uniform pairing order with zero center of mass momentum for the Cooper pair q=0 ($\Delta_0$).  \textbf{b} The intra-branch pairing gives rise to the pairing states with the finite center of mass momentum q=$\pm Q$ ($\Delta_{\pm Q}$). Both pairings can be equal-spin and opposite-spin states. Here s=$\pm$ is the branch index, due to the spin-orbit coupling (SOC), each branch has both spin-up and spin-down components.
			\label{fig:fs}}
	\end{center}
	\vskip-0.5cm
\end{figure}

This effective Hamiltonian can describe the embedded quantum structures~\cite{yi_2021}, such as the domain wall~\cite{liu_2023} and atomic line defects~\cite{chen_2020} in monolayer Fe(Te,Se) grown on SrTiO$_3$.
The effective attraction $V_1$ and $V_2$ can be different since the system already breaks the spin rotation symmetry. Here, the interaction term breaks SU(2) spin rotation symmetry down to U(1) about $z$ axis when $V_1\neq V_2$, while the SOC term breaks the SU(2) symmetry down to U(1) about a general axis determined by $\alpha_R$ and $\alpha_D$, therefore the Hamiltonian fully breaks the spin rotation symmetry.
The connection between the model Hamiltonian and the 1D quantum structure in monolayer Fe(Te,Se) stems from their symmetry properties. In particular, the inversion symmetry breaking due to the substrate induces the Rashba SOC and the mirror symmetry breaking along the domain wall or line defects introduces the Dresselhaus SOC. As a result, the spin-rotation symmetry is completely broken, which we show plays a crucial role in the realization of the PDW state. 
The two attraction terms can be decomposed into equal-spin and opposite-spin pairing channels so that the full Hamiltonian can be written as
\begin{equation}
 \begin{split}
	\hat{H}=&\sum_{k,\sigma\sigma^{\prime}} c_{k\sigma}^{\dagger} \left[ h_0(k) \right]_{\sigma\sigma^{\prime}} c_{k\sigma^{\prime}}
 \\
 &-N_{c}V_{1}\sum_{q\sigma}\hat{\Delta}_{\parallel,q,\sigma}^{\dagger}\hat{\Delta}_{\parallel,q,\sigma}-N_{c}V_{2}\sum_{q\sigma}\hat{\Delta}_{\perp,q,\sigma}^{\dagger}\hat{\Delta}_{\perp,q,\sigma}
 \end{split}
	\label{eq:htbq}
\end{equation}
where the pairing operators in the two channels are defined as
\begin{equation}
	\begin{cases}
		\hat{\Delta}_{\parallel,q,\sigma}=\frac{1}{N_{c}}\sum_{k} i\sin k c_{-k+\frac{q}{2}\sigma}c_{k+\frac{q}{2}\sigma}\\
		\hat{\Delta}_{\perp,q,\sigma}=\frac{1}{N_{c}}\sum_{k}e^{-ik} c_{-k+\frac{q}{2}\bar{\sigma}} c_{k+\frac{q}{2}\sigma}
	\end{cases}
	\label{eq:op}
\end{equation}
and the non-interacting Hamiltonian has the form
\begin{equation}
    h_0(k)=(-2t\cos{k}-\mu)\sigma_0-2\sin{k}(\alpha_R\sigma_z + \alpha_D\sigma_x) 
    \label{eq:h0}
\end{equation}
Here, $k$ is the momentum, $\mu$ is the chemical potential, 
\{$\sigma_0$, $\sigma_x$, $\sigma_y$, $\sigma_z$\} are the identity matrix, and three Pauli matrices in the spin space and $N_c$ is the number of sites. $h_0(k)$ describes the bare bands spitted by SOC as shown schematically in Fig.~\ref{fig:fs}. \\

\noindent \textbf{Pairing order parameters}\\
Due to the presence of both $\alpha_R$ and $\alpha_D$, the wave function of each branch of the band contains both spin components, therefore, based on the structure of the Fermi surface shown in Fig.~\ref{fig:fs}, the most general mean-field ansatz that respects the time-reversal symmetry can be written as
\begin{equation}
\begin{cases}
	\left\langle \hat{\Delta}_{\parallel,0,\uparrow}\right\rangle =\left\langle \hat{\Delta}_{\parallel,0,\downarrow}\right\rangle^* =\Delta_{\parallel,0} \ , \ 
        \left\langle \hat{\Delta}_{\perp,0,\uparrow}\right\rangle =-\left\langle \hat{\Delta}_{\perp,0,\downarrow}\right\rangle^*=  \Delta_{\perp,0}
 \\
	\left\langle \hat{\Delta}_{\parallel,Q,\uparrow}\right\rangle =\left\langle \hat{\Delta}_{\parallel,-Q,\downarrow}\right\rangle^* =\Delta_{\parallel,Q} \ , \ 
        \left\langle \hat{\Delta}_{\perp,Q,\uparrow}\right\rangle=-\left\langle \hat{\Delta}_{\perp,-Q,\downarrow}\right\rangle^*=\Delta_{\perp,Q}
 \\
	\left\langle \hat{\Delta}_{\parallel,-Q,\uparrow}\right\rangle =\left\langle \hat{\Delta}_{\parallel,Q,\downarrow}\right\rangle^* =\Delta_{\parallel,-Q} \ , \ 
        \left\langle \hat{\Delta}_{\perp,-Q,\uparrow}\right\rangle=-\left\langle \hat{\Delta}_{\perp,Q,\downarrow}\right\rangle^*=\Delta_{\perp,-Q}
\end{cases}
\label{eq:ansatz}
\end{equation}
Here, the electrons belonging to the opposite branches can form Cooper pairs with zero center of mass momentum. Since each branch of the bands contains both spin components, the pairing can be either equal-spin pairing $\Delta_{\parallel,0}$ or opposite-spin pairing $\Delta_{\perp,0}$.
While the equal-spin pairing must be spin-triplet, the opposite-spin pairing is in general mixed-parity pairing where the mixture of the spin-singlet and triplet pairing is controlled by the phase of the order parameter $\Delta_{\perp,0}$, with phase $0$ corresponding to the $s$-wave pairing and phase $\pi$/2 corresponding to the $p$-wave pairing and other phase values corresponding to the mixture of both.
On the other hand, the electrons belonging to the same branch can pair up with the nonzero center of mass momentum $Q$ and $-Q$, which is determined by the SOC as
	$Q=2\arctan\left(\frac{\sqrt{\alpha_{R}^{2}+\alpha_{D}^{2}}}{t}\right)$.
Due to the same reason, the finite momentum intra-branch pairing can also be in the equal-spin ($\Delta_{\parallel,Q}$, $\Delta_{\parallel,-Q}$) and opposite-spin pairing channels ($\Delta_{\perp,Q}$, $\Delta_{\perp,-Q}$).
The mean-field equations are then solved with a fixed chemical potential $\mu$ in the folded Brillouin Zone (BZ) for the commensurate $Q$. Assuming $Q=\frac{\pi}{D}$, the BZ is folded into $\frac{1}{2D}$ of the original BZ. \\

\begin{figure}
	\begin{center}
        \includegraphics[width=3.4in]{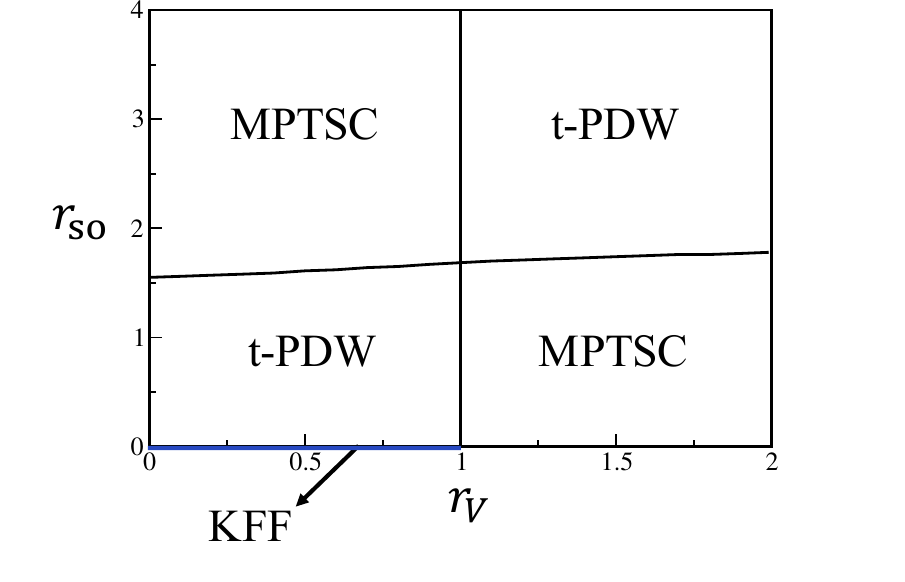}
  \caption{\textbf{Phase diagrams close to half filling.} The spin-triplet pair density wave (t-PDW) and the mixed-parity topological superconductor (MPTSC) states are realized by tuning the two ratios $r_{\rm so}$ and $r_V$ in the mean-field calculations, where the chemical potential $\mu=-0.4$ close to half filling. The Kramers Fulde-Ferrell (KFF) state is realized in the $x$-axis with $r_{\rm so}=0$ and $r_V<1$ represented by the bold blue line.
			\label{fig:pd_mu}}
	\end{center}
	\vskip-0.5cm
\end{figure}

\noindent \textbf{Phase diagram}\\
We first choose the parameters as $t=1$, $\mu=-0.4$, $\sqrt{V_1^2+V_2^2}=1.5$,  $\sqrt{\alpha_{R}^{2}+\alpha_{D}^{2}}=\tan\left(\pi/8\right)$, so that $Q=\frac{\pi}{4}$ and tune the ratio $r_{\rm so}$ and $r_V$. The ground state should be determined by comparing the energy of possible mean-field solutions and we find that the two most competitive states are the t-PDW state and the uniform MP state. 
The obtained phase diagram as functions of $r_{\rm so}$ and $r_V$ is shown in Fig.~\ref{fig:pd_mu}, where two phase boundaries separate the t-PDW and MP states and cross at a single point.
The phase diagram has several features.
First of all, when $r_{\rm so}=0$, the ground state is always the Kramers Fulde-Ferrell (KFF) state for $r_V<1$ until it reaches the transition point at $r_V=1$ and the system transits to the MP state through a first-order phase transition which is studied in the previous work~\cite{KFF}.
The KFF can be viewed as a special limit of the t-PDW state, where the Cooper pairs carry opposite momenta in the opposite spin channels. It has no spatial modulation in the local density of states, although its pairing function carries the finite center of mass momenta. This can be proved by connecting the KFF state to a uniform $p$-wave superconductor through a gauge transformation~\cite{KFF}. However, when $r_{\rm so}$ is nonzero, the ground state becomes the pure t-PDW state for $r_{\rm so}<1$ since the spatial modulation of the order parameter can no longer be gauged away.
As $r_{\rm so}$ further increases, the system goes through a first-order phase transition and the ground state becomes the uniform MP state.
Intriguingly, the phase structure is inverted for the case where $r_V>1$. The ground state remains the uniform MP state when nonzero  $r_{\rm so}$ is introduced until the ratio is big enough to induce a first-order transition from the MP state to the t-PDW state.
While the horizontal transition line is determined numerically, the vertical transition line can be determined analytically and is verified by numerical calculations. 

The vertical transition line corresponds to the case with $r_V=1$, where the interaction part of the Hamiltonian has the full SU(2) spin rotation symmetry. Then for the 1D system, the SOC reduces this symmetry to the U(1) symmetry corresponding to the spin rotation around the special axis with unit vector $\hat{n}_{\alpha}=(\alpha_D,0,\alpha_R)/\sqrt{\alpha_{R}^{2}+\alpha_{D}^{2}}$. Then one can always choose the spin quantization axis along the axis $\hat{n}_{\alpha}$ and the Hamiltonian thus becomes
\begin{equation}
	\begin{split}
		\hat{H}&=\sum_{i\sigma\sigma^{\prime}} c_{i\sigma}^{\dagger} \left(-t\sigma_0+i\sqrt{\alpha_R^2+\alpha_D^2}\sigma_z  \right)_{\sigma\sigma^{\prime}} c_{i+1\sigma^{\prime}} +h.c.
		\\
		&-V_1\sum_{i\sigma\sigma^{\prime}}n_{i\sigma}n_{i+1\sigma^{\prime}}
	\end{split}
\label{eq:bd}
\end{equation}
which corresponds to the transition points at $(r_{\rm so}=0, r_V=1)$ in the phase diagram, where the KFF state and MP state are degenerate. Therefore, we can conclude that along the line with $r_V=1$, the finite momentum pairing state and the uniform MP state are always degenerate, and the system will go through a first-order phase transition by crossing this line.

To understand the horizontal phase transition line, we can start with the left bottom of the phase diagram in Fig.~\ref{fig:pd_mu}.
In this region where $r_V<1$, the equal-spin pairing state is favored over the opposite-spin pairing state.
Moreover, since $\alpha_R$ dominates over $\alpha_D$ in this region, each branch of the Fermi surface mainly consists of the same spin polarization, which means the equal-spin pairing is mainly contributed by the intra-branch pairing as shown schematically in Fig.~\ref{fig:fs}\textbf{b}, leading to the t-PDW state. 
Next, as $\alpha_D$ increases, which enhances the mixture of the two spin components within each branch of the Fermi surface, more contribution of the equal-spin pairing comes from the inter-branch pairing with zero center-of-mass momentum as depicted in Fig.~\ref{fig:fs}\textbf{a}. Therefore, the system is driven into a uniform MP state. 
Contrarily, if we start from the right bottom of the phase diagram, where $r_V>1$ and $\alpha_D$ dominates over $\alpha_R$, the opposite-spin pairing is favored which mainly comes from the inter-branch pairing giving rise to the uniform MP state. Then increasing $\alpha_D$ drives the system into the t-PDW state by pushing more opposite-spin pairing into the intra-branch pairing channel. A formal derivation of the energetics of the t-PDW state in the two limits of $r_{\rm so}\rightarrow 0, \pi/2$ is provided in the Methods section. \\

\begin{figure}
	\begin{center}
	\includegraphics[width=3.4in]{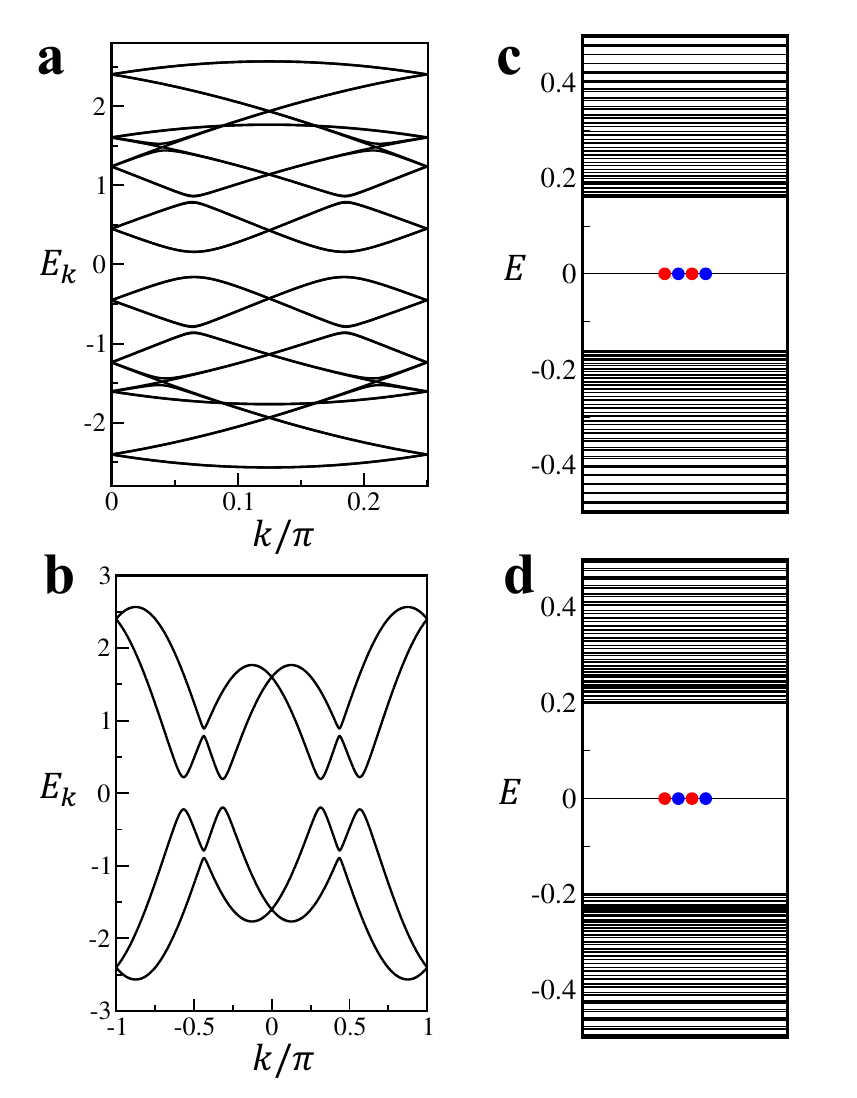}
  \caption{\textbf{Properties of typical spin-triplet pair density wave (t-PDW) and mixed-parity (MP) states.} \textbf{a},\textbf{b} The band dispersion of t-PDW (\textbf{a}) and MP (\textbf{b}) states, where that of the t-PDW state is plotted in the folded Brillouin zone. \textbf{c},\textbf{d} The energy spectrum obtained in a finite-size system with L=320, hosting four zero-energy MZMs denoted by blue and red dots.
  The parameters used are $r_{\rm so}$=1.4 and $r_V$=0.3 for the t-PDW state and $r_{\rm so}$=3.0 and $r_V$=0.3 for the MP state with $\mu=-0.4$ and all other parameters mentioned in the main text.
			\label{fig:topo}}
	\end{center}
	\vskip-0.5cm
\end{figure}

\noindent \textbf{Real space modulations of the $d$-vector in t-PDW}\\
We next study the obtained pairing order parameters of all the states in the phase diagram, which helps to visualize the spatial modulation of the t-PDW state. 
As mentioned above, the real space pairing order parameters can be compactly expressed by Eq.~\ref{eq:dv} as 
\[\Delta_{i,\sigma\sigma^{\prime}}=\left[ (\psi_i \sigma_0 + \mathbf{d}_i\cdot\pmb{\sigma}) i\sigma_y \right]_{\sigma\sigma^{\prime}}
\]
Therefore, both the $s$-wave and $p$-wave components can be expressed as functions of the order parameters defined in Eq.~\ref{eq:ansatz}.
The mean-field t-PDW solution has four finite real order parameters as $(\Delta_{\parallel,Q}, \Delta_{\parallel,-Q}, \Delta_{\perp,Q}=-\Delta_{\perp,-Q})$ leading to the $p$-wave PDW states with the spatial modulated $d$-vector as
\begin{equation}
	\begin{cases}
			d_{i,x}=-i(\Delta_{\parallel,Q}-\Delta_{\parallel,-Q})\sin\left[Q(r_i+\frac{1}{2})\right] \\
			d_{i,y}=-i(\Delta_{\parallel,Q}+\Delta_{\parallel,-Q})\cos\left[Q(r_i+\frac{1}{2})\right] \\
			d_{i,z}=-2i\Delta_{\perp,Q}\sin\left[Q(r_i+\frac{1}{2})\right]
	\end{cases}
 \label{eq:dvect}
\end{equation}
The real space $d$-vector is pure imaginary due to the time-reversal symmetry.
The spatial modulation of a typical $d$-vector of the t-PDW state constructed from the pairing order parameters determined self-consistently at $r_{\rm so}=1.4$ and $r_V=0.3$ is shown schematically in Fig.~\ref{fig:pd_gen}, where the $d$-vector revolves around the origin in an elliptical orbit in 3D space within a period of the PDW state. 

On the other hand, the KFF state at $r_{\rm so}=0$ and $r_V<1$ has only one finite order parameter $\Delta_{\parallel,Q}$, corresponding to the $d$-vector 
\[
\mathbf{d}_i=-i|\Delta_{\parallel,Q}|\left\{\sin\left[Q(r_i+\frac{1}{2})+\phi_{\parallel}\right] , \cos\left[Q(r_i+\frac{1}{2})+\phi_{\parallel}\right], 0 \right\}
\]
with $\phi_{\parallel}=\arg(\Delta_{\parallel,Q})$.
Then the revolving orbit of the $d$-vector around the origin becomes circular in the $xy$ plane.
The uniform MP solution has two finite order parameters with a complex $\Delta_{\perp,0}$ and pure imaginary $\Delta_{\parallel,0}$, which leads to the uniform pairing order parameter as 
$\psi_{i}= -\text{Re} [\Delta_{\perp,0}]$
and $\mathbf{d}_i=-\left\{\Delta_{\parallel,0} , 0, i\text{Im}[\Delta_{\perp,0}] \right\}$, which is shown schematically in the lower pannel of Fig.~\ref{fig:pd_gen}. \\

\noindent \textbf{Topological properties}\\
The band dispersion of both t-PDW and MP states are shown in Fig.~\ref{fig:topo}\textbf{a},\textbf{b}, both of which are fully gapped. In order to determine the topological properties of these states, we calculate the end states of finite-size systems in each phase. As shown in Fig.~\ref{fig:topo}\textbf{c},\textbf{d}, both phases host two pairs of Majorana zero modes (MZMs) at each end of the chain, indicating the nontrivial topological properties. To fully capture the topological properties of both states, we calculate their $Z_2$ topological invariant, ${\cal N}$. While a standard method exists for calculating the $Z_2$ invariant for MP states~\cite{qi_2010}, applying it to the translation symmetry breaking PDW states proves nontrivial. Therefore, we generalize this method to calculate the $Z_2$ invariant for the PDW state (see Methods). We apply this method to the t-PDW states and find that the $Z_2$ topological invariant ${\cal N}=-1$, consistent with the presence of topological boundary states, specifically the two pairs of MZMs shown in Fig.~\ref{fig:topo}\textbf{b}.
Since the time-reversal symmetry is preserved for both phases, they can be classified as time-reversal invariant (TRI) TSC.
The nontrivial topological property of the t-PDW state is easy to understand since it is a pure $p$-wave spin-triplet pairing state with $d$-vector given by Eq.~\ref{eq:dvect} and the spatial modulation of the $d$-vector does not change its topological property.
In contrast, the MP state has both $s$-wave and $p$-wave pairing components and can be either a TSC or a topologically trivial superconductor~\cite{yi_2021}. For the chemical potential close to half-filling with $\mu=-0.4$, the MP state is always topological, which is labeled as MPTSC in the phase diagram of Fig.~\ref{fig:pd_mu}. However, a topological phase transition can occur when the chemical potential is tuned away from the half-filling, which is studied below. \\

\begin{figure}
	\begin{center}
        \includegraphics[width=3.4in]{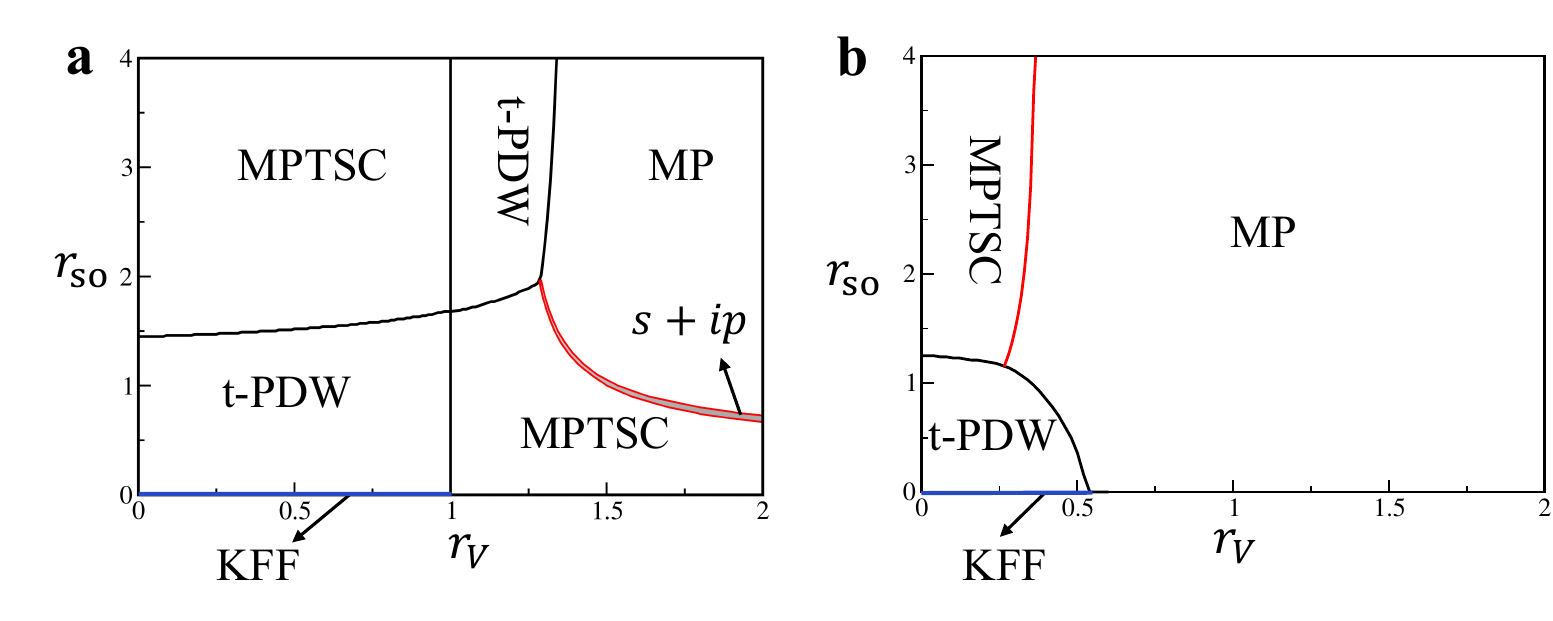}
  \caption{\textbf{Phase diagrams away from half filling.} The chemical potential is set as $\mu$=-1.4 (\textbf{a}) and  $\mu=-1.8$ (\textbf{b}) with all the other parameters unchanged. The Kramers Fulde-Ferrell (KFF) states are realized along the $x$-axis, indicated by the bold blue lines. The narrow phase region shaded in grey and bounded by two nearby red lines in \textbf{a} corresponds to the time-reversal symmetry-breaking $s$+i$p$ state, which emerges near the transition between two time-reversal invariant mixed-parity (MP) states without gap closing~\cite{Ezawa_2013,wang_2016}. The single red line in \textbf{b} represents the continuous phase transition between the topological superconductor (TSC) and trivial superconductor with mixed parity. 
			\label{fig:pd_mu12}}
	\end{center}
	\vskip-0.5cm
\end{figure}

\noindent \textbf{Chemical potential effect}\\
Finally, we study the effect of the chemical potential on the phase diagram. The calculation above is done for $\mu=-0.4$ which is around the half-filling of the chain.
We find that if the chemical potential is tuned away from the half-filling, the region for the t-PDW phase shrinks as shown in Fig.~\ref{fig:pd_mu12}\textbf{a} for $\mu=-1.4$. For $\mu$ close enough to the band top or bottom, the phase diagram changes dramatically, where the t-PDW state can only be realized for small values of $r_V$ and $r_{\rm so}$ which is shown in Fig.~\ref{fig:pd_mu12}\textbf{b} with $\mu=-1.8$. 

The dramatic change of the phase diagram in Fig.~\ref{fig:pd_mu12}\textbf{b} can be understood as follows. 
One important change in the phase diagrams is the absence of the vertical phase boundary with $r_V=1$, where the finite momentum pairing state and uniform MP state are degenerate.
We can first analyze these two degenerate states along this phase boundary. As shown in Eq.~\ref{eq:bd}, along the whole phase boundary, the Hamiltonian is identical to the one with $\alpha_D=0$ and $V_1=V_2$ by choosing a proper spin quantization axis due to the remaining U(1) symmetry.   
Then we can perform a gauge transformation $c_{i
\sigma}\rightarrow e^{i\sigma\theta_{\alpha}r_i}d_{i\sigma}$, and Eq.~\ref{eq:bd} becomes
\begin{equation}
	\begin{split}
		\hat{H}_d&=-t_{\alpha}\sum_{i\sigma} d_{i\sigma}^{\dagger}d_{i+1\sigma} +h.c.
		-V_1\sum_{i\sigma\sigma^{\prime}}n^d_{i\sigma}n^d_{i+1\sigma^{\prime}}
	\end{split}
\end{equation}
where $t_{\alpha}=\sqrt{t^2+\alpha_R^2+\alpha_D^2}$, $\theta_{\alpha}=\arctan\left(\frac{\sqrt{\alpha_R^{2}+\alpha_D^{2}}}{t}\right)=\frac{Q}{2}$ and $n^d_{i\sigma}=d_{i\sigma}^{\dagger}d_{i\sigma}$.
Since the SOC is gauged away in the transformed Hamiltonian, which respects the full spin rotation and inversion symmetry, the ground state of such a system is either the uniform $s$-wave pairing state or the $p$-wave state with its $d$-vector pointing in any direction, depending on the chemical potential. For $\mu$ around half filling, the Fermi points $k_f$ are round $\pm\frac{\pi}{2}$ where the pairing function for the $p$-wave state $\Delta_p(k)\propto\sin k$ gains more energy than that of the $s$-wave state $\Delta_s(k)\propto\cos k$, making the $p$-wave pairing state as the ground state. On the other hand, if $\mu$ is sufficiently tuned away from half-filling, the $s$-wave pairing will take over as $k_f$ approaches $\pm\pi$, where $\Delta_s(k)\propto\cos k$ gains more energy.

We next undo the gauge transformation to see what these states are in the original basis.
For the uniform $p$-wave state in the transformed basis, the pairing order parameter in real space can be expressed as
\begin{equation}
	\tilde{\Delta}_{i,\sigma\sigma^{\prime}}=\left\langle d_{i\sigma}d_{i+1\sigma^{\prime}}\right\rangle=\left[ (\mathbf{\tilde{d}}\cdot\pmb{\sigma}) i\sigma_y \right]_{\sigma\sigma^{\prime}}
\end{equation}
with $\mathbf{\tilde{d}}$ a pure imaginary constant vector.
Then in the original basis, we have 
\begin{equation}
\begin{split}
\Delta_{i,\sigma\sigma^{\prime}}&=\left\langle c_{i\sigma}c_{i+1\sigma^{\prime}}\right\rangle=\left\langle d_{i\sigma}d_{i+1\sigma^{\prime}}\right\rangle e^{i\theta_{\alpha}(\sigma r_i+\sigma^{\prime} r_{i+1})}
\\
&=\begin{cases}
	\tilde{\Delta}_{i,\sigma\sigma} e^{i\sigma Q(r_i+\frac{1}{2})}, \ \ \ \text{if}\  \sigma^{\prime}=\sigma \\
    \tilde{\Delta}_{i,\sigma\bar\sigma} e^{-i\sigma \frac{Q}{2}}, \ \ \ \text{if}\  \sigma^{\prime}=\bar\sigma 
\end{cases}
\end{split}
\end{equation}
Therefore, if $\mathbf{\tilde{d}}$ is in the $x$-$y$ plane corresponding to the equal-spin pairing,
$\Delta_{i,\sigma\sigma}=\tilde{\Delta}_{i,\sigma\sigma} e^{i\sigma Q(r_i+\frac{1}{2})}=(-\tilde{d}_x\sigma+i\tilde{d}_y)e^{i\sigma Q(r_i+\frac{1}{2})}$.
This pairing order parameter corresponds to the $d$-vector in the original basis as
\[
\mathbf{d}_i=i|\tilde{d}|\left\{\sin\left[Q(r_i+\frac{1}{2})+\phi_{\tilde{d}}\right] , \cos\left[Q(r_i+\frac{1}{2})+\phi_{\tilde{d}}\right], 0 \right\}
\]
with $\phi_{\tilde{d}}=\arctan\left( \frac{\tilde{d}_y}{\tilde{d}_x} \right)$,
which is exactly the $d$-vector for the KFF state.
On the other hand, if $\mathbf{\tilde{d}}$ is along the $z$ direction corresponding to opposite-spin pairing, 
$\Delta_{i,\sigma\bar\sigma}=\tilde{\Delta}_{i,\sigma\bar\sigma} e^{-i\sigma \frac{Q}{2}}=-\tilde{d}_z e^{-i\sigma \frac{Q}{2}}$. This corresponds to the
uniform pairing order parameter with
$\psi=-|\tilde{d}_z| \sin(\frac{Q}{2})  $ 
and $\mathbf{d}=-i\left\{0 , 0, |\tilde{d}_z| \cos(\frac{Q}{2}) \right \}$, which describes the MP state. Therefore, the finite momentum pairing state and the uniform pairing state along the vertical phase boundary are mapped onto the $p$-wave pairing state with in-plane and out-of-plane $d$-vectors through a gauge transformation, which are always degenerate in energy. This means as long as $\mu$ is close to half-filling, the vertical phase boundary always exists as shown in Fig.~\ref{fig:pd_mu} and Fig.~\ref{fig:pd_mu12}\textbf{a}.
On the other hand, if $\mu$ is close to the band top or bottom which leads to $s$-wave paring ground state in the transformed basis with pairing order parameter $\tilde{\Delta}_{i,\sigma\sigma^{\prime}}=\left\langle d_{i\sigma}d_{i+1\sigma^{\prime}}\right\rangle=\left( \tilde{\psi} i\sigma_y \right)_{\sigma\sigma^{\prime}}$,
the states in the original basis become
$\Delta_{i,\sigma\bar\sigma}=\tilde{\Delta}_{i,\sigma\bar\sigma} e^{-i\sigma \frac{Q}{2}}=\sigma\tilde{\psi} e^{-i\sigma \frac{Q}{2}}$, corresponding to the uniform MP state with pairing order parameters described by 
$\psi=\tilde{\psi} \cos(\frac{Q}{2})$
and $\mathbf{d}=-i\left\{0 , 0, \tilde{\psi} \sin(\frac{Q}{2}) \right \}$. 
In this case, there is no vertical phase boundary in the phase diagram, and only a small enough $r_V$ can drive the system into a t-PDW state. Therefore, most parts of the phase diagram are occupied by the uniform MP state with the t-PDW state shrinking to the left bottom as shown in Fig.~\ref{fig:pd_mu12}\textbf{b}. \\

\noindent {\large\textbf{Conclusions}}\\
In this work, we systematically study the effective 1D model with general SOC and nearest-neighbor attraction. We find that the triplet pairing PDW state whose $d$-vector modulates in real space can be realized in this microscopic model.
Moreover, the parameter region to realize such a PDW state is large when the chemical potential $\mu$ is close to half-filling and shrinks when $\mu$ approaches the band top or bottom, which is accompanied by the topological phase transition within the MP state.
This work introduces a simple and concrete theoretical model to realize the PDW states.
It also provides a possible explanation for the PDW state reported along the 1D domain wall in monolayer iron-based superconductor Fe(Te,Se) grown on STO.
We hope it will stimulate the exploration of the PDW state in embedded quantum structures of high-temperature superconductors. \\

\noindent {\large\textbf{Methods}}\\
\noindent \textbf{Two pairing channels}\\
\noindent The two attraction terms in Eq.~\ref{eq:ham} can be written as 
\begin{equation}
	\begin{split}
		\hat{H}_{I}&=-\frac{V_1}{N_{c}}\sum_{k,k^{\prime},q,\sigma}\sin k\sin k^{\prime}c_{k+\frac{q}{2}\sigma}^{\dagger}c_{-k+\frac{q}{2}\sigma}^{\dagger}c_{-k^{\prime}+\frac{q}{2}\sigma}c_{k^{\prime}+\frac{q}{2}\sigma}
		\\
		&-\frac{V_2}{N_{c}}\sum_{k,k^{\prime},q,\sigma}e^{i(k-k^{\prime})}c_{k+\frac{q}{2}\sigma}^{\dagger}c_{-k+\frac{q}{2}\bar{\sigma}}^{\dagger}c_{-k^{\prime}+\frac{q}{2}\bar{\sigma}}c_{k^{\prime}+\frac{q}{2}\sigma}
	\end{split}
	\label{eq:Hamil}
\end{equation}
which can be further decomposed into equal-spin and opposite-spin pairing channels as
\begin{equation}
	\hat{H}_{I}=-N_{c}V_{1}\sum_{q\sigma}\hat{\Delta}_{\parallel,q,\sigma}^{\dagger}\hat{\Delta}_{\parallel,q,\sigma}-N_{c}V_{2}\sum_{q\sigma}\hat{\Delta}_{\perp,q,\sigma}^{\dagger}\hat{\Delta}_{\perp,q,\sigma}
\end{equation}
where the pairing operators in the two channels are defined in Eq.~\ref{eq:op}, leading to the full Hamiltonian defined in Eq.~\ref{eq:htbq}. \\

\noindent \textbf{Fermi surface of the bare bands}\\
\noindent The non-interacting Hamiltonian $h_0(k)$ in Eq.~\ref{eq:h0} has eigenvalues correspond to the bare band dispersion as
\begin{equation}
	\varepsilon_{k,s}^{0}= -2t_{\alpha}\cos(k-s\theta_{\alpha})-\mu, \ \  s=\pm
\end{equation}
with $t_{\alpha}=\sqrt{t^{2}+\alpha_R^{2}+\alpha_D^{2}}$ and $\theta_{\alpha}=\arctan\left(\frac{\sqrt{\alpha_R^{2}+\alpha_D^{2}}}{t}\right)=\frac{Q}{2}$.
The Fermi points are thus determined as 
$k_{f,s,\pm}=s\theta_{\alpha}\pm\arccos\left(-\frac{\mu}{2t_{\alpha}}\right)$
where $s=\pm$ corresponds to the two branches of the Fermi surfaces as shown schematically in Fig.~\ref{fig:fs}. \\

\noindent \textbf{Order parameters and mean-field Hamiltonian}\\
After the mean-field decoupling using the ansatz defined by Eq.~\ref{eq:ansatz}, the mean-field Hamiltonian can be written as
\begin{widetext}
\begin{equation}
	\begin{split}
	\hat{H}_{MF}-\mu\hat{N}&=\sum_{k,\sigma,\sigma^{\prime}}c_{k\sigma}^{\dagger}\left[(-2t\cos{k}-\mu)\sigma_0-2\sin{k}(\alpha_R\sigma_z + \alpha_D\sigma_x) \right]_{\sigma\sigma^{\prime}}c_{k\sigma^{\prime}}
	-V_{1}\sum_{k}(-i\sin k) \left( \Delta_{\parallel,0}c_{k\uparrow}^{\dagger}c_{-k\uparrow}^{\dagger}+\Delta_{\parallel,0}^*c_{k\downarrow}^{\dagger}c_{-k\downarrow}^{\dagger} \right)
	\\
	&-V_{1}\sum_{k}(-i\sin k)\left(\Delta_{\parallel,Q}c_{k+\frac{Q}{2}\uparrow}^{\dagger}c_{-k+\frac{Q}{2}\uparrow}^{\dagger} + \Delta_{\parallel,Q}^* c_{k-\frac{Q}{2}\downarrow}^{\dagger}c_{-k-\frac{Q}{2}\downarrow}^{\dagger}+ \Delta_{\parallel,-Q}c_{k-\frac{Q}{2}\uparrow}^{\dagger}c_{-k-\frac{Q}{2}\uparrow}^{\dagger} + \Delta_{\parallel,-Q}^* c_{k+\frac{Q}{2}\downarrow}^{\dagger}c_{-k+\frac{Q}{2}\downarrow}^{\dagger} \right)
	\\
	&-V_{2}\sum_{k}\left( \Delta_{\perp,Q} e^{ik} +\Delta_{\perp,-Q}^* e^{-ik} \right) c_{k+\frac{Q}{2}\uparrow}^{\dagger}c_{-k+\frac{Q}{2}\downarrow}^{\dagger}
	-V_{2}\sum_{k}\left( \Delta_{\perp,-Q} e^{ik} +\Delta_{\perp,Q}^* e^{-ik} \right) c_{k-\frac{Q}{2}\uparrow}^{\dagger}c_{-k-\frac{Q}{2}\downarrow}^{\dagger}
	\\
	&-V_{2}\sum_{k}\left( \Delta_{\perp,0}e^{ik} + \Delta_{\perp,0}^* e^{-ik}\right) c_{k\uparrow}^{\dagger}c_{-k\downarrow}^{\dagger}  
	+h.c.+2N_{c}[V_{1}(|\Delta_{\parallel,0}|^{2}+|\Delta_{\parallel,Q}|^{2}+|\Delta_{\parallel,-Q}|^{2})+V_{2}(|\Delta_{\perp,0}|^{2}+|\Delta_{\perp,Q}|^{2}+|\Delta_{\perp,-Q}|^{2})]
	\end{split}
    \label{eq:Hmf}
\end{equation}
\end{widetext}
The order parameters are determined self-consistently for a fixed chemical potential $\mu$ with the self-consistent equations
\begin{equation}
	\begin{cases}
		\Delta_{\parallel,0}=\frac{1}{N_{c}}\sum_{k}i\sin k\left\langle c_{-k\uparrow}c_{k\uparrow}\right\rangle \\
		\Delta_{\parallel,Q}=\frac{1}{N_{c}}\sum_{k}i\sin k\left\langle c_{-k+\frac{Q}{2}\uparrow}c_{k+\frac{Q}{2}\uparrow}\right\rangle \\
		\Delta_{\parallel,-Q}=\frac{1}{N_{c}}\sum_{k}i\sin k\left\langle c_{-k-\frac{Q}{2}\uparrow}c_{k-\frac{Q}{2}\uparrow}\right\rangle \\
		\Delta_{\perp,0}=\frac{1}{N_{c}}\sum_{k}e^{-ik}\left\langle c_{-k\downarrow}c_{k\uparrow}\right\rangle \\
    	\Delta_{\perp,Q}=\frac{1}{N_{c}}\sum_{k}e^{-ik}\left\langle c_{-k+\frac{Q}{2}\downarrow}c_{k+\frac{Q}{2}\uparrow}\right\rangle\\
    	\Delta_{\perp,-Q}=\frac{1}{N_{c}}\sum_{k}e^{-ik}\left\langle c_{-k-\frac{Q}{2}\downarrow}c_{k-\frac{Q}{2}\uparrow}\right\rangle
	\end{cases}
    \label{eq:self}
\end{equation}
The self-consistent equations can be solved in the folded Brillouin Zone (BZ) for the commensurate Q, assuming Q=$\frac{\pi}{D}$, the BZ is folded into $\frac{1}{2D}$ of the original BZ. Then the mean-field Hamiltonian should be written in the Nambu basis  $\psi_{k}^{\dagger}=\left(c_{k+(n-1)Q\uparrow}^{\dagger},c_{k+(n-1)Q\downarrow}^{\dagger},c_{-k-(n-1)Q\uparrow},c_{-k-(n-1)Q\downarrow}\right)$, with $n\in[1,2D]$, as
\begin{equation}
	\begin{split}
	\hat{H}_{MF}-\mu\hat{N}&=\frac{1}{4D}\sum_{k}\psi_{k}^{\dagger}h_{k}\psi_{k}
	+2N_{c}[V_{1}(|\Delta_{\parallel,0}|^{2}+|\Delta_{\parallel,Q}|^{2}+|\Delta_{\parallel,-Q}|^{2})
	\\
	&+V_{2}(|\Delta_{\perp,0}|^{2}+|\Delta_{\perp,Q}|^{2}+|\Delta_{\perp,-Q}|^{2})]-\mu N_{c}
	\end{split}
\end{equation}
with $h_{k}$ being the Hamiltonian matrix of size 8D. Here, $h_{k}$ can be written with the block form as
\begin{equation}
	h_{k}=\left[\begin{array}{cc}
		h_{t}(k,Q) & h_{\Delta}(k,Q)\\
		h_{\Delta}^{\dagger}(k,Q) & -h_{t}^{*}(-k,-Q)
	\end{array}\right]
\end{equation}
where the non-zero elements of the matrices are, 
\begin{widetext}
\begin{equation}
	h_{t}(k,Q)_{\{2n-1:2n,2n-1:2n\}}=h_{0}\left(k+(n-1)Q\right).
\end{equation}
\begin{equation}
	h_{\Delta}(k,Q)_{\{2n-1:2n,2n-1:2n\}}=\left[\begin{array}{cc}
		2i V_1 \Delta_{\parallel,0} \sin [k+(n-1)Q]  & -V_2 \left(\Delta_{\perp,0}e^{i[k+(n-1)Q]}+\Delta_{\perp,0}^*e^{-i[k+(n-1)Q]}\right)\\
		V_2 \left(\Delta_{\perp,0}e^{-i[k+(n-1)Q]}+\Delta_{\perp,0}^*e^{i[k+(n-1)Q]}\right) & 2i V_1 \Delta_{\parallel,0}^* \sin [k+(n-1)Q]
	\end{array}\right]
\end{equation}
\begin{equation}
	h_{\Delta}(k,Q)_{\{\left(2n+1\right)^{\prime}:\left(2n+2\right)^{\prime},2n-1:2n\}}=\left[\begin{array}{cc}
			2i V_1 \Delta_{\parallel,Q}\sin[k+(n-\frac{1}{2})Q] & -V_2 \left(\Delta_{\perp,Q}e^{i[k+(n-1/2)Q]}+\Delta_{\perp,-Q}^*e^{-i[k+(n-1/2)Q]}\right)\\
		V_2 \left(\Delta_{\perp,Q}e^{-i[k+(n-1/2)Q]}+\Delta_{\perp,-Q}^*e^{i[k+(n-1/2)Q]}\right) & 2i V_1 \Delta_{\parallel,-Q}^*\sin[k+(n-\frac{1}{2})Q]
	\end{array}\right]
\end{equation}
\begin{equation}
	h_{\Delta}(k,Q)_{\{2n-1:2n,\left(2n+1\right)^{\prime}:\left(2n+2\right)^{\prime}\}}=\left[\begin{array}{cc}
		2i V_1 \Delta_{\parallel,-Q}\sin[k+(n-\frac{1}{2})Q] & -V_2 \left(\Delta_{\perp,Q}^*e^{-i[k+(n-1/2)Q]}+\Delta_{\perp,-Q}e^{i[k+(n-1/2)Q]}\right)\\
		V_2 \left(\Delta_{\perp,Q}^*e^{i[k+(n-1/2)Q]}+\Delta_{\perp,-Q}e^{-i[k+(n-1/2)Q]}\right) & 2i V_1 \Delta_{\parallel,Q}^*\sin[k+(n-\frac{1}{2})Q]
	\end{array}\right]
\end{equation}
\end{widetext}
for $n\ensuremath{\in[1,2D]}$ and $\left(2n+1\right)^{\prime}=\left(2n\right)\mod(4D)+1$ and $\left(2n+2\right)^{\prime}=\left(2n+1\right)\mod(4D)+1$, so that all the indices are within the range of [1,4D].
The self-consistent loop begins with an initial guess for the order parameters $\left(\Delta_{\parallel,0}, \Delta_{\parallel,Q}, \Delta_{\parallel,-Q}, \Delta_{\perp,0}, \Delta_{\perp,Q}, \Delta_{\perp,-Q}\right)$. Using these parameters, the meanfield Hamiltonian (Eq.\ref{eq:Hmf}) is diagonalized in the folded BZ. The resulting ground-state expectation values are then computed using Eq.\ref{eq:self} to obtain an updated set of order parameters. These updated parameters are linearly mixed with the previous set and used to construct the mean-field Hamiltonian for the next iteration. The process is repeated until convergence, defined as the numerical equality of the old and updated order parameters within a specified tolerance. \\

\noindent \textbf{Relation between the $d$-vectors and the order parameters}\\
The real space pairing order parameters can be calculated as
\begin{equation}
	\begin{cases}
		\Delta_{i,\uparrow\uparrow}=\left\langle c_{i\uparrow}c_{i+1\uparrow}\right\rangle = \Delta_{\parallel,0} + \Delta_{\parallel,Q} e^{iQ(r_i+\frac{1}{2})} + \Delta_{\parallel,-Q} e^{-iQ(r_i+\frac{1}{2})} \\
		\Delta_{i,\downarrow\downarrow}=\left\langle c_{i\downarrow}c_{i+1\downarrow}\right\rangle = \Delta_{\parallel,0}^* + \Delta_{\parallel,Q}^* e^{-iQ(r_i+\frac{1}{2})} + \Delta_{\parallel,-Q}^* e^{iQ(r_i+\frac{1}{2})} \\
		\Delta_{i,\uparrow\downarrow}=\left\langle c_{i\uparrow}c_{i+1\downarrow}\right\rangle = -\Delta_{\perp,0} - \Delta_{\perp,Q} e^{iQ(r_i+\frac{1}{2})} - \Delta_{\perp,-Q} e^{-iQ(r_i+\frac{1}{2})} \\
		\Delta_{i,\downarrow\uparrow}=\left\langle c_{i\downarrow}c_{i+1\uparrow}\right\rangle = \Delta_{\perp,0}^* + \Delta_{\perp,Q}^* e^{-iQ(r_i+\frac{1}{2})} + \Delta_{\perp,-Q}^* e^{iQ(r_i+\frac{1}{2})}
	\end{cases}
\end{equation}
Therefore, we can compactly express the pairing order parameters according to Eq.~\ref{eq:dv} as
\[
	\Delta_{i,\sigma\sigma^{\prime}}=\left[ (\psi_i \sigma_0 + \mathbf{d}_i\cdot\pmb{\sigma}) i\sigma_y \right]_{\sigma\sigma^{\prime}}
\]
where $\psi_i$ and $\mathbf{d}_i$ correspond to the $s$-wave and $p$-wave components of the pairing order parameters and can be expressed as
\begin{equation}
	\begin{cases}
		\begin{split}
			\psi_{i}&=-\frac{1}{2}[ (\Delta_{\perp,0}+\Delta_{\perp,0}^*) + (\Delta_{\perp,Q}+\Delta_{\perp,-Q}^*)e^{iQ(r_i+\frac{1}{2})} 
			\\ &+(\Delta_{\perp,-Q}+\Delta_{\perp,Q}^*)e^{-iQ(r_i+\frac{1}{2})} ]
		\end{split}  \\
	    \begin{split}
		d_{i,x}&=-\frac{1}{2}[ (\Delta_{\parallel,0}-\Delta_{\parallel,0}^*) +(\Delta_{\parallel,Q}-\Delta_{\parallel,-Q}^*)e^{iQ(r_i+\frac{1}{2})} \\
		&+(\Delta_{\parallel,-Q}-\Delta_{\parallel,Q}^*)e^{-iQ(r_i+\frac{1}{2})} ]
		\end{split}  \\
	    \begin{split}
		d_{i,y}&=-\frac{i}{2}[ (\Delta_{\parallel,0}+\Delta_{\parallel,0}^*) +(\Delta_{\parallel,Q}+\Delta_{\parallel,-Q}^*)e^{iQ(r_i+\frac{1}{2})} \\
		&+(\Delta_{\parallel,-Q}+\Delta_{\parallel,Q}^*)e^{-iQ(r_i+\frac{1}{2})} ]
		\end{split} \\
	    \begin{split}
		d_{i,z}&=-\frac{1}{2}[ (\Delta_{\perp,0}-\Delta_{\perp,0}^*) +(\Delta_{\perp,Q}-\Delta_{\perp,-Q}^*)e^{iQ(r_i+\frac{1}{2})} \\
		&+(\Delta_{\perp,-Q}-\Delta_{\perp,Q}^*)e^{-iQ(r_i+\frac{1}{2})} ]
		\end{split}
	\end{cases}
\end{equation}
For the time-reversal symmetric states, $\psi_i$ should be real, and $\mathbf{d}_i$ should be imaginary. \\

\noindent \textbf{Competition between finite momentum pairing and zero momentum pairing states}\\
Since the finite (zero) momentum pairing states come from the intra-branch (inter-branch) pairing as shown in Fig.~\ref{fig:fs} and finite $r_{\rm so}$ mixes the spin components within the branch, we can perform a change of basis as $ \left(\begin{array}{c}
c_{i\uparrow}^{\prime}\\
c_{i\downarrow}^{\prime}
\end{array}\right)=e^{\frac{i}{2}\theta\sigma_{y}}\left(\begin{array}{c}
c_{i\uparrow}\\
c_{i\downarrow}
\end{array}\right) $ with $\theta=\arctan(r_{\rm so})$,
which aligns the spin quantization axis with the spin polarizations of the two branches of the non-interacting band.
In this new basis, the non-interacting Hamiltonian becomes
\begin{equation}
    h_0^{\prime}(k)=-2t\cos k\sigma_{0}-2\sqrt{\alpha_{R}^{2}+\alpha_{D}^{2}}\sin k\sigma_{z}
\end{equation}
where the spin and the branch indices are locked so that the Fermi surface instability leads to finite momentum equal spin pairing and zero momentum opposite spin pairing.
However, this makes the interacting part more complicated which reads
\begin{equation}
\begin{split}
    H_I^{\prime}&=-V_{1}^{\prime}\sum_{i\sigma}n_{i\sigma}^{\prime}n_{i+1\sigma}^{\prime} - V_{2}^{\prime} \sum_{i\sigma}n_{i\sigma}^{\prime}n_{i+1\bar{\sigma}}^{\prime}
    \\
    &+V_{\theta 1}^{\prime}\sum_{i\sigma} \left(c_{i\sigma}^{\prime\dagger}c_{i+1\sigma}^{\prime\dagger}c_{i+1\bar{\sigma}}^{\prime}c_{i\bar{\sigma}}^{\prime}+c_{i\sigma}^{\prime\dagger}c_{i+1\bar{\sigma}}^{\prime\dagger}c_{i+1\sigma}^{\prime}c_{i\bar{\sigma}}^{\prime}\right)
    \\
    &+V_{\theta 2}^{\prime}\sum_{i\sigma} \left[\sigma c_{i\sigma}^{\prime\dagger}c_{i+1\sigma}^{\prime\dagger}\left(c_{i+1\sigma}^{\prime}c_{i\bar{\sigma}}^{\prime}+c_{i+1\bar{\sigma}}^{\prime}c_{i\sigma}^{\prime}\right)+h.c.\right]
\end{split}
\label{eq:HIp}
\end{equation}
where
\begin{equation}
	\begin{cases}
		V_{1}^{\prime}=V_{1}-\frac{1}{2}(V_{1}-V_{2})\sin^{2}\theta \\
		V_{2}^{\prime}=V_{2}+\frac{1}{2}(V_{1}-V_{2})\sin^{2}\theta \\
		V_{\theta 1}^{\prime}=-\frac{1}{2}(V_{1}-V_{2})\sin^{2}\theta \\
		V_{\theta 2}^{\prime}=\frac{1}{4}(V_{1}-V_{2})\sin2\theta
	\end{cases}
\end{equation}
It is obvious that when $V_1=V_2$ ($r_V=1$), the interacting part is invariant under this U(1) spin rotation, and the finite momentum pairing state i.e., the KFF state and the MP state are degenerate as demonstrated in the previous section.
When $r_V\neq 1$, this degeneracy is lifted and the ground state is determined by $r_V$ and $\theta=\arctan(r_{\rm so})$, leading to the phase diagram in Fig.~\ref{fig:pd_mu}.

We can understand how the finite momentum pairing state is favored in the two limits with $\theta\rightarrow 0$ and $\theta\rightarrow \pi/2$.
When $\theta\rightarrow 0$, the two density-density term in Eq.~\ref{eq:HIp} dominate and leads to the KFF state with order parameter $\Delta_{\parallel,Q}^{\prime}$ when $r_V<1$ since $V_{1}^{\prime}-V_{2}^{\prime}=(V_{1}-V_{2})\cos^2\theta>0$. The next order effect comes from the $V_{\theta 1}^{\prime}$ term in Eq.~\ref{eq:HIp}, which can be written in the mean-field level as
\begin{equation}
   H_{I,\theta 1}^{\prime}=V_{\theta 1}^{\prime}\left[\text{Re}(\Delta_{\parallel,Q}^{\prime} \Delta_{\parallel,-Q}^{\prime})-\text{Re}(\Delta_{\perp,0}^{\prime 2})\right] \ .
   \label{eq:Htheta1}
\end{equation}
This term induces a small amount of KFF state $\Delta_{\parallel,-Q}^{\prime}$ leading to the t-PDW state.
The $V_{\theta 2}^{\prime}$ term induces a spin-triplet pairing with opposite spin components.
When $\theta\rightarrow \pi/2$, we have $V_{1}^{\prime} \approx V_{2}^{\prime}$ and $V_{\theta 2}^{\prime} \approx 0$, which means it is $H_{I,\theta 1}^{\prime}$ that lifts the degeneracy of the finite momentum and zero momentum pairing state and determines the ground state.
As demonstrated in the previous section, the zero momentum MP state in this limit has the order parameter 
$\Delta_{\perp,0}^{\prime}\propto i e^{-iQ/2}$ leading to $\text{Re}(\Delta_{\perp,0}^{\prime 2})\propto -\cos Q <0$. Considering the relation $|\Delta_{\parallel,Q}^{\prime}| \approx |\Delta_{\perp,0}^{\prime}| > |\Delta_{\parallel,-Q}^{\prime}|$ in this limit, we can conclude that the t-PDW state gains more energy when $V_{\theta 1}^{\prime}>0$ i.e., $r_V>1$ and the MP state becomes the ground state when $V_{\theta 1}^{\prime}<0$ i.e., $r_V<1$, which explains the main phase diagram shown in Fig.~\ref{fig:pd_mu}. \\

\noindent \textbf{Topological phase transition within the MP state away from half-filling}\\
Another feature of the phase diagram when the chemical potential is tuned away from the half-filling is the existence of an extra topological phase transition within the MP state with the phase boundary represented by the red lines in Fig.~\ref{fig:pd_mu12}\textbf{a} and Fig.~\ref{fig:pd_mu12}\textbf{b}. These phase boundaries separate the TSC and the trivial superconductor, whose topological property can be determined by a $Z_2$ invariant $\mathcal{N}$~\cite{qi_2010}, which was used to study the topological phase diagram in the atomic line defect~\cite{yi_2021}.

\begin{figure}
	\begin{center}
	\includegraphics[width=3.4in]{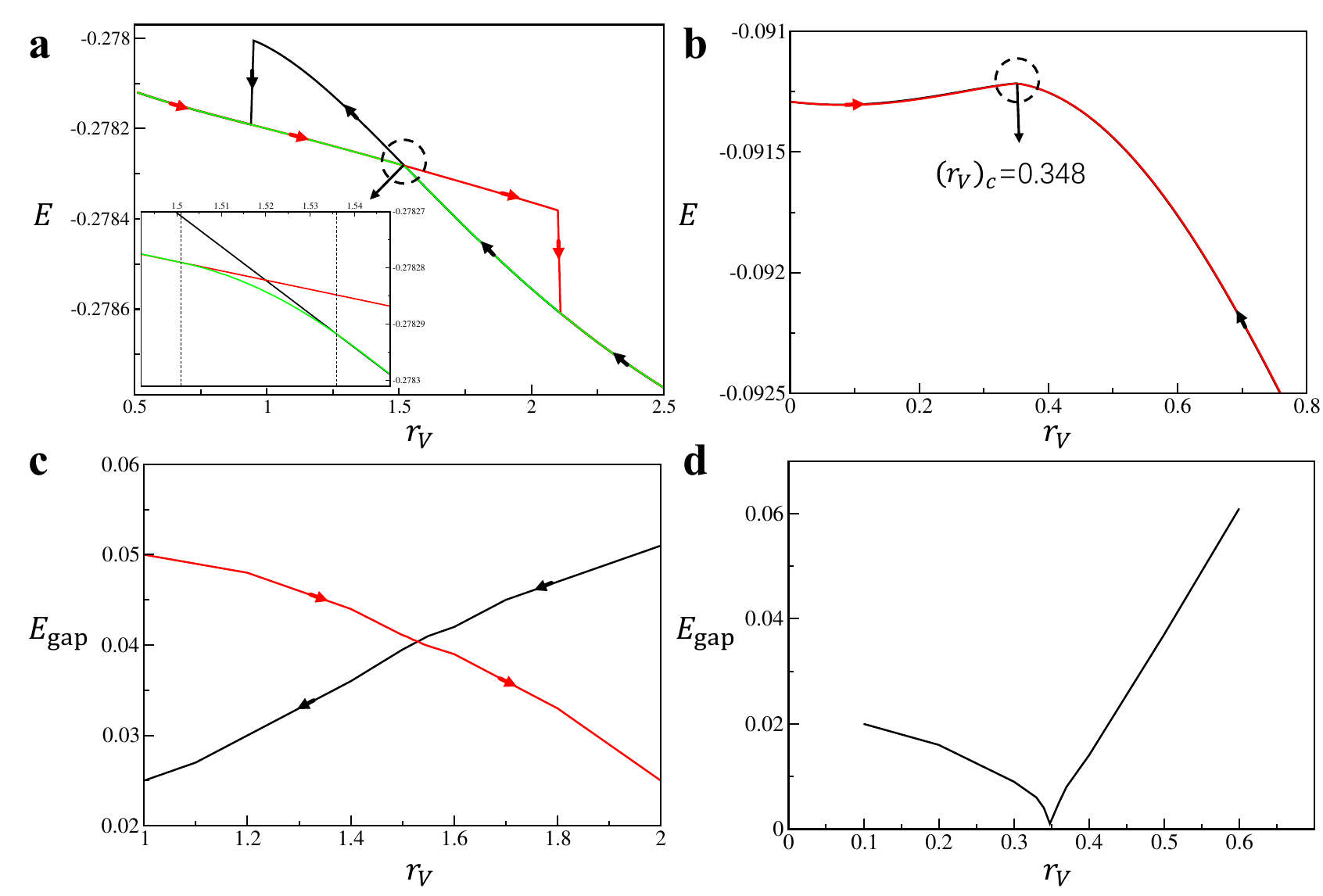}\caption{\textbf{Topological phase transitions within the mixed-parity (MP) states.} \textbf{a}, \textbf{b} The evolution of the total energy across the phase boundary between the topological trivial and non-trivial MP states with the fixed value of $r_{\rm so}=1.0$ and  $\mu$=-1.4 for \textbf{a} and $r_{\rm so}=3.0$ and $\mu$=-1.8 for \textbf{b}. Here, the black (red) line corresponds to tuning $r_V$ from big (small) to small (big) values indicated by the black (red) arrows, showing the hysteresis behavior in \textbf{a} only. The green line in \textbf{a} represents the energy of the time-reversal symmetry-breaking (TRSB) states across the phase boundary. The inset shows a zoomed-in region around the transition points where the two MP states are degenerate, indicating that the TRSB state can become the ground state by mixing the two MP states with a nonzero relative phase near the degeneracy point. The two vertical dashed lines indicate the range where the TRSB state becomes the ground state for a fixed value of $r_{so}$. \textbf{c}, \textbf{d} The evolution of the extracted band gap of the states across the corresponding phase boundary described in \textbf{a} and \textbf{b}. The band gap never closes for the case with $\mu=-1.4$ (\textbf{c}) and closes and reopens across the phase boundary with the critical value $r_V$=0.348 for the case with $\mu=-1.8$ (\textbf{d}).
			\label{fig:pt}}
	\end{center}
	\vskip-0.5cm
\end{figure}

We carefully study the phase transitions for these two cases.
We find that for $\mu=-1.4$ where the vertical phase boundary mentioned above still exists, the topological phase transition is first-order.
To see it more transparently, we fix the ratio $r_{\rm so}=1.0$ and tune the parameter $r_V$ through the phase boundary. We find the total energy has a hysteresis behavior as the parameter $r_V$ is tuned from big to small values and the small to big values respectively which is shown in Fig.~\ref{fig:pt}\textbf{a}, indicating the first-order feature of the transition. Moreover, the band structure of the TSC and trivial SC state are almost identical close to the transition point and the band gap of the system never closes across the phase transition as shown in Fig.~\ref{fig:pt}\textbf{c}.
Across the topological phase transitions without band gap closure~\cite{Ezawa_2013}, the time-reversal symmetry can be spontaneously broken by mixing the degenerate topological trivial and non-trivial MP states~\cite{wang_2016}. We perform the meanfield calculation including TRSB orders and find that such solutions indeed exist. Near the topological phase transition points, where the two MP states are degenerate in energy, these TRSB states become energetically favored over the TRI ones as shown in the inset of Fig.~\ref{fig:pt}\textbf{a}.
These states mix the $s-$ and $p-$ wave pairing with a finite phase difference, giving rise to the $s$+i$p$ states, as depicted by the narrow shaded grey region bounded by the red lines in Fig.~\ref{fig:pd_mu12}\textbf{a}. 
Away from the degenerate points, the $s$+i$p$ states become less favored and the ground states revert to the TRI MP states. We thus conclude that the TRSB $s$+i$p$ states emerge in a narrow region near the phase transition and provide an intriguing pathway across the topological phase transition between two TRI MP states.

As the chemical potential is further tuned towards the band bottom with $\mu=-1.8$, besides the dramatic change of the phase diagram discussed above, the topological phase transition within the MP state also switches from first-order to continuous. 
As an example, we again fix the ratio $r_{\rm so}=3.0$ and tune the parameter $r_V$ through the phase boundary.
In this case, the total energy does not show any hysteresis behavior as shown in Fig.~\ref{fig:pt}\textbf{b}. We further calculate the band dispersion of the state as the parameters are tuned across the phase boundary, which shows a typical gap close-and-reopen behavior across the topological phase transition as shown in Fig.~\ref{fig:pt}\textbf{d}. \\

\noindent \textbf{Topological invariant for the MP and PDW states}\\
Since the 1D model respects both the time-reversal and particle-hole symmetry, it belongs to class $DIII$~\cite{altland_1997} whose topological property can be described by a $Z_2$ invariant $\mathcal{N}$~\cite{qi_2010}. 
A simple expression for computing this invariant involves the expectation value of the time-reversed pairing functions for each band $n$ at momentum $k$:
\begin{eqnarray}
	\delta_{nk}=\langle n,k| {\cal T} h_{\Delta}^{\dagger}(k) |n,k \rangle
    \label{eq:Z2_1}
\end{eqnarray}
where $|n,k \rangle$ denotes the eigenstate of the normal state Hamiltonian $h_N(k)$, ${\cal T}$  is the time-reversal matrix satisfying ${\cal T}^\dagger h_N(k){\cal T}=h_N^T(-k)$ and $h_{\Delta}(k)$ is pairing matrix. In the weak pairing limit, the topological invariant ${\cal N}$ in 1D is given by the product
\begin{eqnarray}
	{\cal N}=\Pi_{s} [{\rm sgn}(\delta_s)]
    \label{eq:Z2_2}
\end{eqnarray}
where $s$ runs over all Fermi points of all bands between $0$ and $\pi$.

For the MP state where only the two order parameters $\Delta_{\parallel,0}$ and $\Delta_{\perp,0}$ are finite, the pairing matrix becomes
\begin{equation}
\begin{split}
   h_{\Delta}(k)&=V_1 \left( \text{Re}[\Delta_{\parallel,0}]i\sigma_0 - \text{Im}[\Delta_{\parallel,0}]\sigma_z \right) \sin{k} 
   \\
   &+ V_2 \left( \text{Im}[\Delta_{\perp,0}]\sigma_x\sin{k} -\text{Re}[\Delta_{\perp,0}]i\sigma_y\cos{k} \right),   
\end{split}
\end{equation}
the time-reversal matrix ${\cal T}=-i\sigma_y$ and $h_N(k)$ is $h_0(k)$ defined in Eq.~\ref{eq:h0}. Then Eqs.~\ref{eq:Z2_1},\ref{eq:Z2_2} can be directly used to calculate the $Z_2$ invariant ${\cal N}$, which leads to the topological phase transition shown in Fig.~\ref{fig:pd_mu12}.

For the PDW state with $Q=\frac{\pi}{D}$, where the finite order parameters are $(\Delta_{\parallel,Q}, \Delta_{\parallel,-Q}, \Delta_{\perp,Q}, \Delta_{\perp,-Q})$, the above method can be generalized to calculate the $Z_2$ invariant ${\cal N}$.
In order to express the time-reversal matrix more conveniently, we choose the basis as $\phi_k^\dagger=\left( c_{k-\frac{2D-1}{2}Q\uparrow}^{\dagger}, c_{k-\frac{2D-1}{2}Q\downarrow}^{\dagger}, c_{k-\frac{2D-3}{2}Q\uparrow}^{\dagger}, c_{k-\frac{2D-3}{2}Q\downarrow}^{\dagger}, \cdots , c_{k+\frac{2D-1}{2}Q\uparrow}^{\dagger}, c_{k+\frac{2D-1}{2}Q\downarrow}^{\dagger} \right)$
so that the Nambu basis is defined as $\Psi_k^{\dagger}=\left( \phi_k^{\dagger}, \phi_{-k}^{T} \right)$ and $k$ is defined in the folded BZ, i.e. $k\in[-\frac{Q}{2},\frac{Q}{2}]$.
In this basis, the time-reversal matrix becomes ${\cal T}=J_{2D}\otimes(-i\sigma_y)$, where $J_{2D}$ is the counteridentity matrix whose elements are all equal to zero except those on the antidiagonal, which are all equal to 1 with dimension $2D$ and the normal state Hamiltonian $h_N(k)$ becomes block-diagonal which can be expressed as
\begin{equation}
    h_N(k)=\bigoplus_{i=1}^{2D} h_{0}\left(k+\frac{2i-2D-1}{2}Q\right).
\end{equation}
Moreover, the pairing matrix $h_{\Delta}(k)$ becomes a matrix of size 4D, whose block form can be expressed as
\begin{equation}
	\begin{cases}
		h_{\Delta}(k)_{n,2D-n}=h_{p,-}\left(k+\frac{2n-2D-1}{2}Q\right), n\in [1,2D-1] \\
		h_{\Delta}(k)_{n,2D+2-n}=h_{p,+}\left(k+\frac{2n-2D-1}{2}Q\right), n\in [1,2D-1] \\
		h_{\Delta}(k)_{2D,2D}=h_{p,-}\left(k+\frac{2D-1}{2}Q\right) \\
		h_{\Delta}(k)_{1,1}=h_{p,+}\left(k+\frac{1-2D}{2}Q\right)
	\end{cases}
    \label{eq:hp}
\end{equation}
where,
\begin{widetext}
\begin{equation}
	h_{p,-}(k)=\left[\begin{array}{cc}
			2i V_1 \Delta_{\parallel,-Q}\sin(k+\frac{1}{2}Q) & -V_2 \left[\Delta_{\perp,-Q}e^{i(k+\frac{1}{2}Q)}+\Delta_{\perp,Q}^*e^{-i(k+\frac{1}{2}Q)}\right]\\
		V_2 \left[\Delta_{\perp,-Q}e^{-i(k+\frac{1}{2}Q)}+\Delta_{\perp,Q}^*e^{i(k+\frac{1}{2}Q)}\right] & 2i V_1 \Delta_{\parallel,Q}^*\sin(k+\frac{1}{2}Q)
	\end{array}\right]
\end{equation}
\begin{equation}
	h_{p,+}(k)=\left[\begin{array}{cc}
		2i V_1 \Delta_{\parallel,Q}\sin(k-\frac{1}{2}Q) & -V_2 \left[\Delta_{\perp,-Q}^*e^{-i(k-\frac{1}{2}Q)}+\Delta_{\perp,Q}e^{i(k-\frac{1}{2}Q)}\right]\\
		V_2 \left[\Delta_{\perp,-Q}^*e^{i(k-\frac{1}{2}Q)}+\Delta_{\perp,Q}e^{-i(k-\frac{1}{2}Q)}\right] & 2i V_1 \Delta_{\parallel,-Q}^*\sin(k-\frac{1}{2}Q)
	\end{array}\right]
\end{equation}
\end{widetext}
Then the $Z_2$ invariant can be calculated using Eq.~\ref{eq:Z2_1},\ref{eq:Z2_2}, with $s$ running over all Fermi points of all bands between $0$ and $\frac{Q}{2}$.
For the t-PDW state, the $Z_2$ invariant ${\cal N}$ is always -1 indicating that the t-PDW state is indeed a TRI TSC, which is consistent with the two pairs of MZMs found in the finite length chain. In this specific case, the eigenstates of the normal Hamiltonian are doubly degenerate, we thus introduce a small perturbation proportional to ${\cal T} h_{\Delta}^{\dagger}(k)$ to lift the degeneracy and the resulting eigenstates are used to calculate the $Z_2$ invariant.  \\

\noindent {\large\textbf{Data availability}}\\
The data that support the findings of this study are available from the corresponding authors upon request. \\

\noindent {\large\textbf{Code availability}}\\
The source code of the mean-field calculation is available from the corresponding author on reasonable request. \\

\noindent {\large\textbf{References}}
\bibliography{reference}

\noindent {\large\textbf{Acknowledgments}} \\
YZ is supported in part by the Shanghai Science and Technology Innovation Action Plan (Grant No. 24LZ1400800) and National Natural Science Foundation of China (NSFC) Grants No. 12274279.
ZW is supported by the U.S. Department of Energy, Basic Energy Sciences, Grant No. DE FG02-99ER45747. \\

\noindent {\large\textbf{Author contributions}} \\
Y.Z. and Z.W. contributed to all aspects of this work and wrote the paper. \\

\noindent {\large\textbf{Competing interests}} \\
The authors declare no competing interests. \\

\end{document}